\documentclass[eqsecnum,twocolumn,aps,epsf,superscriptaddress]{revtex4}
\usepackage{graphicx}
\usepackage{color}

\begin{document}
\draft

\title{Strong electron correlation behind the superconductivity \\ in Ce-free and Ce-underdoped high-$T_{\rm c}$ T'-cuprates}

\author{T. Adachi}
\thanks{Corresponding author: t-adachi@sophia.ac.jp}
\affiliation{Department of Engineering and Applied Sciences, Sophia University, 7-1 Kioi-cho, Chiyoda-ku, Tokyo 102-8554, Japan}

\author{A. Takahashi}
\author{K. M. Suzuki}
\thanks{Present address: Institute for Materials Research, Tohoku University, 2-1-1 Katahira, Aoba-ku, Sendai 980-8577, Japan}
\author{M. A. Baqiya}
\author{T. Konno}
\author{T. Takamatsu}
\author{M. Kato}
\affiliation{Department of Applied Physics, Graduate School of Engineering, Tohoku University, 6-6-05 Aoba, Aramaki, Aoba-ku, Sendai 980-8579, Japan}

\author{I. Watanabe}
\affiliation{Advanced Meson Science Laboratory, Nishina Center for Accelerator-Based Science, The Institute of Physical and Chemical Research (RIKEN), 2-1 Hirosawa, Wako 351-0198, Japan}

\author{A. Koda}
\author{M. Miyazaki}
\thanks{Present address: Applied Physics Research Unit, Muroran Institute of Technology, 27-1 Mizumoto-cho, Muroran 050-8585, Japan}
\author{R. Kadono}
\affiliation{Muon Science Laboratory, Institute of Materials Structure Science, High Energy Accelerator Research Organization (KEK), 1-1 Oho, Tsukuba 305-0801, Japan}

\author{Y. Koike}
\affiliation{Department of Applied Physics, Graduate School of Engineering, Tohoku University, 6-6-05 Aoba, Aramaki, Aoba-ku, Sendai 980-8579, Japan}

\date{\today}

\begin{abstract}
In order to investigate the electronic state of Ce-free and Ce-underdoped high-$T_{\rm c}$ cuprates with the so-called T' structure, we have performed muon-spin-relaxation ($\mu$SR) and specific-heat measurements of Ce-free T'-La$_{1.8}$Eu$_{0.2}$CuO$_{4+\delta}$ (T'-LECO) polycrystals and Ce-underdoped T'-Pr$_{1.3-x}$La$_{0.7}$Ce$_x$CuO$_{4+\delta}$ (T'-PLCCO) single crystals with $x = 0.10$.  
The $\mu$SR spectra of the reduced superconducting samples of both T'-LECO with $T_{\rm c} = 15$ K and T'-PLCCO with $x = 0.10$ and $T_{\rm c} = 27$ K have revealed that a short-range magnetic order coexists with the superconductivity in the ground state. 
The formation of a short-range magnetic order due to a tiny amount of excess oxygen in the reduced superconducting samples strongly suggest that the Ce-free and Ce-underdoped T'-cuprates are regarded as strongly correlated electron systems.

\end{abstract}
\vspace*{2em}
\pacs{PACS numbers: }
\maketitle
\newpage

%*****************************************************************************************
%\section{Introduction}\label{intro}
%*****************************************************************************************
In the history of the research of high-temperature superconductivity, enormous efforts have been paid to the establishment of the phase diagram of both hole-doped and electron-doped cuprates.  
For hole-doped La$_{2-x}$Sr$_x$CuO$_4$ with the K$_2$NiF$_4$ (the so-called T) structure, it is well known that the antiferromagnetic (AF) order in the Mott-insulating state of the parent compound is destroyed by only 2\% doping of holes and superconductivity sets in at $x=0.05-0.06$.
It is also well recognized that the Cu-spin correlation, which is believed to be crucial for the appearance of the superconductivity, is characterized by incommensurate magnetic peaks in the elastic/inelastic neutron scattering~\cite{yamada} relating to the so-called spin-charge stripe order.~\cite{tranquada}
For electron-doped Nd$_{2-x}$Ce$_x$CuO$_4$ (NCCO) with the Nd$_2$CuO$_4$ (the so-called T') structure, it has widely been believed that the AF order is formed in the underdoped regime of $x < 0.14$.~\cite{luke}  
Inelastic neutron-scattering experiments have revealed that the commensurate Cu-spin correlation relating to the simple AF order is developed in the electron-doped T'-cuprates.~\cite{yamada-prl}
One of characteristics of the electron-doped T'-cuprates is that as-grown samples contain excess oxygen more or less at the apical site.
The excess oxygen is understood to induce disorder of the electrostatic potential in the CuO$_2$ plane, leading to the destruction of Cooper pairs.~\cite{xu}  
Therefore, the removal of excess oxygen from as-grown samples is crucial for the appearance of superconductivity and the investigation of intrinsic properties of the T'-cuprates. 

Formerly, a surprising result has come out in thin films of T'-cuprates in which superconductivity appears without doping of electrons.~\cite{tsukada} 
This has been followed by a result using T'-NCCO thin films that the superconducting (SC) transition temperature, $T_{\rm c}$, of $\sim 28$ K at Ce-free $x=0$ decreases monotonically with increasing $x$ and disappears at $x \sim 0.20$.~\cite{matsumoto}
The superconductivity in the parent compounds of the T'-cuprates has also been confirmed using polycrystalline samples.~\cite{asai,takamatsu} 
These findings have been brought about through the removal of excess oxygen from as-grown samples more effectively than that ever reported. 
These results have provided following two messages; (i) the newly proposed phase diagram without the AF phase is completely different from the former one, and (ii) the AF Mott-insulating state might not be in intimate relation with the superconductivity in the T'-cuprates.

Related theoretical works on the electronic state in the parent compounds of the T'-cuprates are roughly sorted by the strength of the electron correlation.
In the case of weak electron correlation, a simple band-metal state has been proposed,~\cite{massidda} while in strong electron correlation, calculations based on the local-density approximation (LDA) combined with the dynamical mean-field theory (DMFT) have revealed the possibility of the closing of the so-called charge-transfer (CT) gap between the upper Hubbard band (UHB) of the Cu $3d_{x^2-y^2}$ orbital and the O $2p_{\sigma}$ band observed in the T-cuprates.~\cite{das,weber}
In the case of moderate electron correlation, the appearance of superconductivity due to doublons and holons has been proposed.~\cite{yokoyama} 
Therefore, in order to understand the novel SC state in the T'-cuprates, it is of significance to investigate the Cu-spin correlation, because localized Cu spins in the lower Hubbard band (LHB) of the Cu $3d_{x^2-y^2}$ orbital are generated and expected to correlate with one another in the case of strong electron correlation.

Recently, we have succeeded in obtaining SC single crystals of Ce-underdoped T'-Pr$_{1.3-x}$La$_{0.7}$Ce$_x$CuO$_{4+\delta}$ (T'-PLCCO) with $x = 0.10$ through the improved reduction annealing.~\cite{adachi} 
The $ab$-plane and $c$-axis electrical resistivities in magnetic fields have revealed that, through the reduction annealing, the strongly localized state of carriers accompanied by a pseudogap due to AF fluctuations in the as-grown crystal changes to the Kondo state without pseudogap in the SC crystal. 
The absence of the AF pseudogap in the SC crystal has also been confirmed by the angle-resolved photoemission spectroscopy (ARPES) measurements.~\cite{horio}
These results can be understood in terms of a band picture based on the strong electron correlation.~\cite{adachi}

In order to investigate the Cu-spin correlation in the T'-cuprates in detail, we have carried out muon-spin-relaxation ($\mu$SR) measurements for both as-grown non-SC and reduced SC samples of Ce-free T'-La$_{1.8}$Eu$_{0.2}$CuO$_{4+\delta}$ (LECO) polycrystals and Ce-underdoped T'-PLCCO ($x=0.10$) single crystals. 
It has been found that the Cu-spin correlation is markedly developed and a short-range magnetic order (SRMO) is formed at low temperatures even in the reduced SC samples of T'-LECO and T'-PLCCO. 
Moreover, the estimation of the magnetic and SC volume fractions from $\mu$SR and specific heat has uncovered the coexistence of a SRMO and superconductivity in the reduced SC samples.
The formation of a SRMO due to a tiny amount of excess oxygen in the reduced SC samples strongly suggests that the Ce-free and Ce-underdoped T'-cuprates are regarded as strongly correlated electron systems.

%*****************************************************************************************
%\section{Experimental details}
%*****************************************************************************************
Polycrystalline samples of T'-LECO were prepared by the low-temperature technique using CaH$_2$ as a reductant and subsequent annealing.~\cite{takamatsu} 
The SC samples were obtained through the reduction annealing at 700$^{\rm o}$C for 24 h in vacuum.
Single crystals of T'-PLCCO with $x = 0.10$ were prepared by the traveling-solvent floating-zone technique.~\cite{adachi} 
The SC crystals were obtained through the improved reduction annealing at 800$^{\rm o}$C for 24 h in vacuum.~\cite{adachi}
From the iodometric titration, the oxygen content was confirmed to be reduced by $0.03\pm0.01$ through the reduction annealing for T'-PLCCO with $x = 0.10$.
The magnetic-susceptibility measurements using a SQUID magnetometer (Quantum Design, MPMS) have revealed that $T_{\rm c}$'s of T'-LECO and T'-PLCCO with $x=0.10$ are 15 K and 27 K, respectively. 

Zero-field (ZF) and longitudinal-field (LF) $\mu$SR measurements of T'-LECO were performed using a continuous positive muon beam and the General-Purpose Spectrometer at the Paul-Scherrer Institute (PSI) in Switzerland. 
ZF- and LF-$\mu$SR measurements of T'-PLCCO were performed using a pulsed positive muon beam at the Material and Life Science Experimental Facility at the J-PARC in Japan.
Specific-heat measurements were carried out in magnetic fields of $0 - 9$ T by the thermal relaxation method (Quantum Design, PPMS).

%*****************************************************************************************
%\section{Results}
%*****************************************************************************************
\begin{figure}[tbp]
\includegraphics[width=1.0\linewidth]{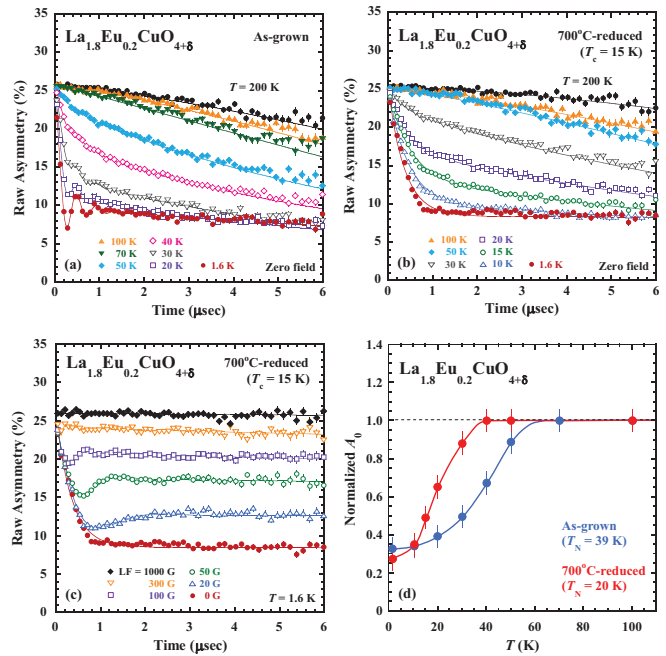}
\caption{(Color online) Zero-field $\mu$SR time spectra of (a) as-grown non-SC and (b) 700$^{\rm o}$C-reduced SC samples of T'-La$_{1.8}$Eu$_{0.2}$CuO$_{4+\delta}$.  Solid lines indicate the best-fit results using the analysis function: $A(t) = A_0 e^{-\lambda_0 t} G_{\rm Z}(\Delta,t) + A_1 e^{-\lambda_1 t} + A_2 e^{-\lambda_2 t} {\rm cos}(\omega t+\phi) + A_{\rm BG}$.  (c) Longitudinal-field $\mu$SR time spectra at 1.6 K of 700$^{\rm o}$C-reduced SC samples of T'-La$_{1.8}$Eu$_{0.2}$CuO$_{4+\delta}$. Solid lines indicate the best-fit results using the analysis function (see, the supplemental materials A).  (d) Temperature dependence of $A_0$ normalized by the value at 200 K in T'-La$_{1.8}$Eu$_{0.2}$CuO$_{4+\delta}$.  Solid lines are to guide the readers' eye.}
\label{fig:f1}
\end{figure}

Figures 1(a) and (b) show the ZF-$\mu$SR time spectra of as-grown non-SC and 700$^{\rm o}$C-reduced SC samples of T'-LECO, respectively.
For the spectra of the as-grown sample, while Gaussian-like slow depolarizations due to the nuclear dipole fields are observed at high temperatures, fast depolarizations and muon-spin precessions appear at low temperatures, indicating the formation of a long-range magnetic order.
Taking a look at the spectrum at 1.6 K, the muon-spin precession is strongly damped, which is similar to that reported in reduced T'-NCCO with $x=0$~\cite{luke} and is different from that observed in T-La$_2$CuO$_4$ with a less damped precession owing to the coherent magnetic order.~\cite{uemura}
These suggest that the long-range AF order in the T'-cuprates is not so coherent throughout the sample.
As for the spectra of the reduced SC sample, although the depolarization becomes fast gradually with decreasing temperature due to the development of the Cu-spin correlation, it is slower than that of the as-grown sample.
At 1.6 K, the spectrum consists of a fast depolarization without precession and a following time-independent behavior above 1 $\mu$s, which is typical of a SRMO state.
This is also corroborated by the LF-$\mu$SR time spectra at 1.6 K shown in Fig. 1(c).  
The spectra shift up in parallel with LF, which is typical of a static magnetically ordered state.
Therefore, it is strongly suggested that the magnetic ground state of the Ce-free reduced SC sample of T'-LECO is a SRMO.

The ZF spectra of T'-LECO were analyzed using the function, $A(t) = A_0 e^{-\lambda_0 t} G_{\rm Z}(\Delta,t) + A_1 e^{-\lambda_1 t} + A_2 e^{-\lambda_2 t} {\rm cos}(\omega t+\phi) + A_{\rm BG}$. 
Here the first and second terms correspond to fast and slowly fluctuating states of Cu spins, respectively. 
The third term represents the muon-spin precession.
The $A_0, A_1, A_2$ and $\lambda_0, \lambda_1$ are initial asymmetries and depolarization rates of each component, respectively.
The $G_{\rm Z}(\Delta,t)$ is the static Kubo-Toyabe function with a half-width of the static internal field at the muon site, $\Delta$, describing the distribution of the nuclear dipole field at the muon site. 
The $\lambda_2, \omega$ and $\phi$ are the damping rate, precession frequency and phase of the precession, respectively.
The $A_{\rm BG}$ is due to the temperature-independent background.
It is found that the magnetic-transition temperature, $T_{\rm N}$, defined at the midpoint of the sharp change of $A_0$ vs. temperature shown in Fig. 1(d) is 39 K for the as-grown sample, which is much lower than $T_{\rm N} = 250$ K for reduced T'-NCCO with $x=0$~\cite{luke} and $T_{\rm N} = 115$ K for reduced T'-La$_2$CuO$_4$.~\cite{hord} 
This is probably due to the small amount of excess oxygen residing in the present as-grown sample of T'-LECO.
As for the reduced SC sample, $T_{\rm N}$ is found to be 20 K.
The internal magnetic field at the muon site, $B_{\rm int}$, is directly estimated from $\omega$ in zero field as $\omega = \gamma_\mu B_{\rm int}$, where $\gamma_\mu$ is the gyromagnetic ratio of muon spin ($\gamma_\mu / 2 \pi = 13.55$ MHz/kOe).
In the SRMO state, it is able to be estimated roughly from the LF-$\mu$SR spectra shown in Fig. 1(c) (see the supplemental material A). 
The estimated values of $B_{\rm int}$ at 1.6 K are 167 G and 39 G for the as-grown and reduced SC samples, respectively. 
To be summarized, through the reduction annealing, the long-range AF order changes to a SRMO with reduced values of $T_{\rm N}$ and $B_{\rm int}$ in T'-LECO.

\begin{figure}[tbp]
\includegraphics[width=1.0\linewidth]{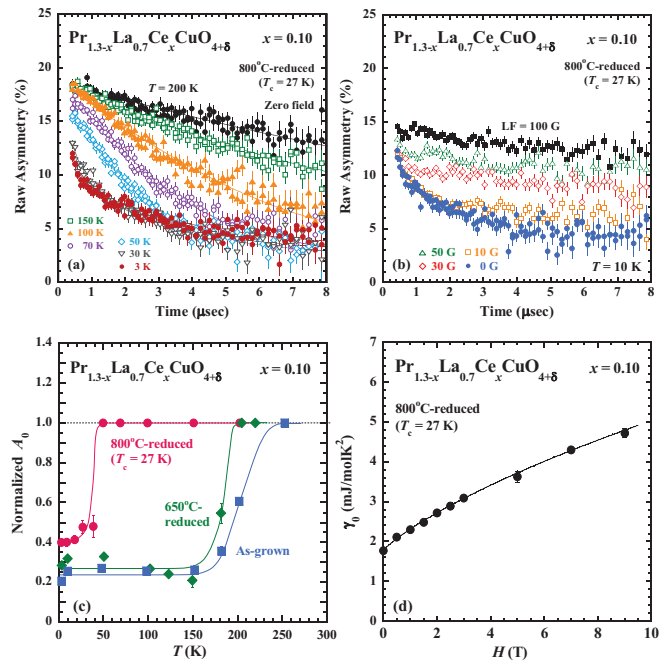}
\caption{(Color online) (a) Zero-field and (b) longitudinal-field $\mu$SR time spectra of the 800$^{\rm o}$C-reduced SC crystal of T'-Pr$_{1.3-x}$La$_{0.7}$Ce$_x$CuO$_{4+\delta}$ with $x=0.10$.  Solid lines in (a) indicate the best-fit results using the analysis function: $A(t) = (A_0 e^{-\lambda_0 t} + A_1 e^{-\lambda_1 t}) e^{-\sigma^2 t^2 / 2} + A_{\rm BG}$.  (c) Temperature dependence of $A_0$ normalized by the value at high temperatures in T'-Pr$_{1.3-x}$La$_{0.7}$Ce$_x$CuO$_{4+\delta}$ with $x=0.10$.  Solid lines are to guide the readers' eye.  (d) Magnetic-field dependence of the residual electronic specific-heat coefficient in the ground state, $\gamma_0$, in the 800$^{\rm o}$C-reduced SC crystal of T'-Pr$_{1.3-x}$La$_{0.7}$Ce$_x$CuO$_{4+\delta}$ with $x=0.10$. The solid line indicate the best-fit result using the function in proportion to $H{\rm ln}H$.~\cite{hlnh}}
\label{fig:f2}
\end{figure}

In T'-PLCCO with $x=0.10$, the ZF-$\mu$SR spectra of the as-grown non-SC crystal exhibit muon-spin precessions at low temperatures (see the supplemental material B). 
As shown in Fig. 2(a), the spectra of the 800$^{\rm o}$C-reduced SC crystal show that the muon-spin depolarization becomes fast with decreasing temperature due to the development of the Cu-spin correlation.
Even at 3 K, however, both of fast and slow depolarizations without muon-spin precession are observed.
Focusing on the spectra in the long-time region, it is found that the asymmetry recovers with decreasing temperature below 50 K, which is analogous to the recovery of the normalized asymmetry to 1/3 usually observed in static magnetically ordered states. 

The magnetic state at low temperatures is uncovered by the LF-$\mu$SR spectra shown in Fig. 2(b).
The spectra change almost in parallel with increasing LF, which is evidence for the formation of a static magnetic state.
Moreover, the spectra exhibit slow depolarizations even in LF, suggesting the existence of fluctuating Cu spins.
Therefore, the LF-$\mu$SR results suggest the coexistence of a static SRMO and dynamically fluctuating Cu spins in the reduced SC crystal of T'-PLCCO with $x=0.10$.

In the analysis of ZF-$\mu$SR spectra of the reduced SC crystal of T'-PLCCO, the following multiple function of exponential- and Gaussian-type was used, considering effects of small Pr$^{3+}$ moments with the Gaussian distribution in a crystal:~\cite{kadono,risdi} $A(t) = (A_0 e^{-\lambda_0 t} + A_1 e^{-\lambda_1 t}) e^{-\sigma^2 t^2 / 2} + A_{\rm BG}$.
The first and second exponential components correspond to dynamically fluctuating Cu spins and a SRMO, respectively. 
The Gaussian component is due to small Pr$^{3+}$ moments and nuclear spins, both of which are dense and randomly distributed. 
Here $\sigma$ is the distribution width of the dipole field at each muon site.
In the analysis of ZF-$\mu$SR spectra of the as-grown non-SC crystal, the equation used for T'-LECO was adopted.~\cite{asfunction}

Figure 2(c) shows the temperature dependence of $A_0$ normalized by the value at high temperatures for T'-PLCCO with $x=0.10$.  
The $T_{\rm N}$ decreases from 203 K of the as-grown crystal to 39 K of the 800$^{\rm o}$C-reduced SC one, indicating that the reduction annealing progressively weakens the static magnetic order. 
Taking a look at the saturated values of $A_0$ at the lowest temperatures, $A_0$ values of the as-grown and 650$^{\rm o}$C-reduced crystals are almost the same as each other, suggesting that the volume fraction of the long-range AF ordered region is almost 100\% in both crystals.
For the 800$^{\rm o}$C-reduced crystal, the saturated value of $A_0$ is apparently larger than those of the other two crystals, and a rough estimation in the 800$^{\rm o}$C-reduced crystal yields that volume fractions of the SRMO and paramagnetic regions are 80\% and 20\%, respectively.

The SC volume fraction is able to be roughly evaluated from the electronic specific heat.
The residual electronic specific-heat coefficient in the SC ground state, $\gamma_0$, which is proportional to the residual density of states at the Fermi level, tends to increase with increasing magnetic field due to the generation of quasiparticles. 
Figure 2(d) shows the magnetic-field dependence of $\gamma_0$ of the 800$^{\rm o}$C-reduced crystal.
In ZF, $\gamma_0$ is finite due to the pair-breaking effect in a dirty $d$-wave superconductor, which is confirmed by $\gamma_0$ in proportion to $H$ln$H$ shown in Fig. 2(d).~\cite{hlnh}
It is found that $\gamma_0$ increases with increasing field and is not saturated at 9 T, indicating that the superconductivity is not completely suppressed at 9 T.
Therefore, it is suggested that the SC volume fraction is at least 62 \% in the 800$^{\rm o}$C-reduced crystal of T'-PLCCO with $x=0.10$ and that the superconductivity spatially coexists with a SRMO.

%*****************************************************************************************
%\section{Discussion}
%*****************************************************************************************

The present ZF- and LF-$\mu$SR of Ce-free T'-LECO and Ce-underdoped T'-PLCCO with $x=0.10$ have revealed that, through the reduction annealing, the long-range AF order in the as-grown samples containing a large amount of excess oxygen changes to a SRMO in the reduced SC samples containing a tiny amount of excess oxygen.  
It would be a reasonable speculation that an ideal T'-cuprate in which the excess oxygen is completely removed exhibits no AF order.
The formation of a SRMO due to a tiny amount of excess oxygen in the reduced SC samples suggests that the Ce-free and Ce-underdoped T'-cuprates are regarded as strongly correlated electron systems.
Therefore, not the proposed simple band-metal state without strong electron correlation~\cite{massidda} but both the doublon-holon model~\cite{yokoyama} and calculations using LDA with DMFT~\cite{das,weber} under the moderate and strong electron correlation, respectively, would be able to explain the superconductivity in the Ce-free and Ce-underdoped T'-cuprates.

\begin{figure}[tbp]
\includegraphics[width=1.0\linewidth]{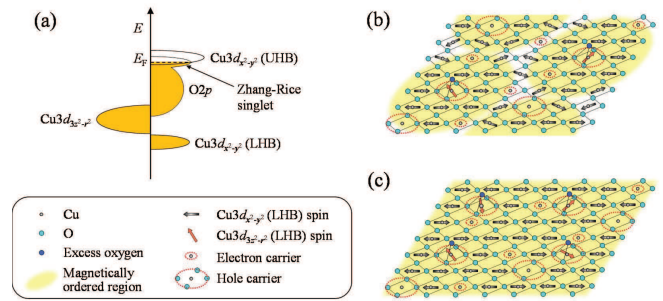}
\caption{(Color online) (a) Band structure of an ideally fully reduced sample of the parent compounds of the T'-cuprates.  (b)(c) Schematic drawings of electron and hole carriers and Cu spins in the CuO$_2$ plane in (b) reduced SC and (c) as-grown non-SC samples of the parent compounds of the T'-cuprates.}
\label{fig:f3}
\end{figure}

Here we discuss how the coexisting state of superconductivity and a SRMO is realized in a reduced SC sample.
As mentioned in the introduction, a band picture based on the strong electron correlation has been proposed, as schematically displayed in Fig. 3(a).~\cite{adachi} 
This is characterized by the collapse of the CT gap due to the reduction of the energy of the Cu $3d_{x^2-y^2}$ orbital owing to the planer square coordination of oxygen around Cu and the finite density of states of the Zhang-Rice singlet band and UHB of the Cu $3d_{x^2-y^2}$ orbital at the Fermi level.
This gives rise to both mobile electrons and holes in the CuO$_2$ plane, leading to the appearance of the superconductivity without nominal doping of carriers by the Ce substitution, as illustrated in Fig. 3(b).
On the other hand, these carriers would be localized around the excess oxygen because of the disorder of the electrostatic potential in the CuO$_2$ plane near the excess oxygen, resulting in the development of the AF correlation of Cu $3d_{x^2-y^2}$ spins in LHB and therefore a SRMO is formed around the excess oxygen.~\cite{kondo} 
In T'-LECO, as shown in Fig. 3(b), the most part of the CuO$_2$ plane would be covered by the SRMO region and the superconductivity would appear around the boundary between SRMO regions.
In T'-PLCCO with $x=0.10$, due to the amount of excess oxygen smaller than that of T'-LECO, the SRMO region decreases in correspondence to the increase in the SC volume fraction.
As for the as-grown non-SC samples of T'-LECO and T'-PLCCO with $x=0.10$ including excess oxygen more than the reduced SC samples, carriers are strongly localized, as confirmed by the electrical-resistivity measurements,~\cite{adachi} and the AF order is long-ranged as illustrated in Fig. 3(c). 
Accordingly, it is suggested that a SRMO is formed around the excess oxygen and the superconductivity appears far from the excess oxygen. 

The magnetic and SC volume fractions estimated have indicated the presence of spatially coexisting SRMO and SC regions in the reduced SC samples.
This kind of spatial coexistence has been observed in so-called multi-layered high-$T_{\rm c}$ cuprates~\cite{mukuda} and the iron-arsenide superconductor LaFeAsO$_{1-x}$F$_x$.~\cite{hiraishi}
In the present reduced SC samples, a part of the SRMO region appears to be spatially overlapped with the SC region, but it is possible that the SRMO and SC regions are phase-separated, because the volume fraction of SRMO might be overestimated due to electronic dipole fields being extended into the paramagnetic SC region. 

Recent ARPES experiments by Horio {\it et al}.~\cite{horio} using our reduced SC crystals of T'-PLCCO with $x=0.10$ have provided an intriguing result in relation to a SRMO.
The AF pseudogap has been found to be closed on the whole Fermi surface, which is contrary to the former ARPES results in the T'-cuprates.~\cite{armitage,matsui}
Considering both results of ARPES without the AF pseudogap and $\mu$SR with a SRMO, it is likely that Cu spins forming a SRMO have tiny magnetic moments. 
That is, the reduction annealing gives rise to not only the change of the long-range AF order to a SRMO but also the reduction of the magnetic moments of Cu spins, which is compatible with the decrease in $B_{\rm int}$ in the present result and with the recent neutron-scattering result in Ce-free T'-Pr$_{1.4}$La$_{0.6}$CuO$_{4+\delta}$ that the intensity of the magnetic excitation spectra significantly decreases through the reduction annealing.~\cite{tsutsumi}

The observation of a SRMO in the reduced SC sample of T'-LECO should be compared with the recent ZF-$\mu$SR result by Kojima {\it et al}.~\cite{kojima} using a reduced SC thin film of Ce-free T'-La$_{1.9}$Y$_{0.1}$CuO$_4$ that the magnetic state depends on the depth from the surface of the thin film.
While a disordered magnetic state is observed near the surface of the thin film, the Cu-spin correlation is rather developed even in the inside of the film.
These results suggest that the Ce-free superconductivity appears in a state where the Cu-spin correlation is developed, which is basically consistent with the present results.

%*****************************************************************************************
%\section{Conclusion}
%*****************************************************************************************
In conclusion, in the T'-cuprates of Ce-free T'-LECO and Ce-underdoped T'-PLCCO with $x=0.10$, the ZF- and LF-$\mu$SR measurements have revealed that a SRMO is formed at low temperatures.
From the estimation of the volume fractions of the SRMO region from $\mu$SR and the SC region from the specific heat, it has been clarified that SRMO spatially coexists with the superconductivity in the reduced SC samples. 
By using our proposed band picture based on the strong electron correlation, it has been suggested that a SRMO is formed around the residual excess oxygen in the reduced SC samples due to the localization of carriers and that the superconductivity appears far from the excess oxygen.
Accordingly, a SRMO brought about with a tiny amount of excess oxygen suggests that the superconductivity in the Ce-free and Ce-underdoped T'-cuprate occurs under the strong electron correlation, which is the same as that in the hole-doped T-cuprates.

%*****************************************************************************************
%\section*{Acknowledgments}
%*****************************************************************************************
Fruitful discussions with A. Fujimori, M. Fujita, M. Horio, M. Ogata, K. Tsutsumi, H. Yokoyama and T. Yoshida are gratefully acknowledged. 
We also thank A. Amato and H. Luetkens at PSI for their technical support in the $\mu$SR measurements.
This work was supported by JSPS KAKENHI Grant Number 23540399 and 23108004 and by Sophia University Special Grant for Academic Research.


\begin{references}

\bibitem{yamada}
K. Yamada, C. H. Lee, K. Kurahashi, J. Wada, S. Wakimoto, S. Ueki, H. Kimura, Y. Endoh, S. Hosoya, G. Shirane, R. J. Birgeneau, M. Greven, M. A. Kastner, and Y. J. Kim,
 Phys. Rev. B {\bf 57}, 6165 (1998).

\bibitem{tranquada}
J. M. Tranquada, B. J. Sternlieb, J. D. Axe, Y. Nakamura, and S. Uchida,
 Nature (London) {\bf 375}, 561 (1995).

\bibitem{luke}
G. M. Luke, L. P. Le, B. J. Sternlieb, Y. J. Uemura, J. H. Brewer, R. Kadono, R. F. Kiefl, S. R. Kreitzman, T. M. Riseman, C. E. Stronach, M. R. Davis, S. Uchida, H. Takagi, Y. Tokura, Y. Hidaka, T. Murakami, J. Gopalakrishnan, A. W. Sleight, M. A. Subramanian, E. A. Early, J. T. Markert, M. B. Maple, and C. L. Seaman,
 Phys. Rev. B {\bf 42}, 7981 (1990).

\bibitem{yamada-prl}
K. Yamada, K. Kurahashi, T. Uefuji, M. Fujita, S. Park, S.-H. Lee, and Y. Endoh,
 Phys. Rev. Lett. {\bf 90}, 137004 (2003).

\bibitem{xu}
X. Q. Xu, S. N. Mao, W. Jiang, J. L. Peng, and R. L. Greene,
 Phys. Rev. B {\bf 53}, 871 (1996).

\bibitem{tsukada}
A. Tsukada, Y. Krockenberger, M. Noda, H. Yamamoto, D. Manske, L. Alff, and M. Naito,
 Solid State Commun. {\bf 133}, 427 (2005).

\bibitem{matsumoto}
O. Matsumoto, A. Utsuki, A. Tsukada, H. Yamamoto, T. Manabe, and M. Naito,
 Physica C {\bf 469}, 924 (2009).

\bibitem{asai}
S. Asai, S. Ueda, and M. Naito,
 Physica C {\bf 471}, 682 (2011).

\bibitem{takamatsu}
T. Takamatsu, M. Kato, T. Noji, and Y. Koike,
 Appl. Phys. Express {\bf 5}, 073101 (2012).

\bibitem{massidda}
S. Massidda, N. Hamada, J. Yu, and A. J. Freeman,
 Physica C {\bf 157} 571 (1989).

\bibitem{das}
H. Das, and T. Saha-Dasgupta,
 Phys. Rev. B {\bf 79} 134522 (2009).

\bibitem{weber}
C. Weber, K. Haule, and G. Kotliar:
 Nature Phys. {\bf 6} 574 (2010).

\bibitem{yokoyama}
H. Yokoyama, M. Ogata, and Y. Tanaka,
 J. Phys. Soc. Jpn. {\bf 75} 114706 (2006).

\bibitem{adachi}
T. Adachi, Y. Mori, A. Takahashi, M. Kato, T. Nishizaki, T. Sasaki, N. Kobayashi, and Y. Koike,
 J. Phys. Soc. Jpn. {\bf 82}, 063713 (2013).

\bibitem{horio}
M. Horio, T. Adachi, Y. Mori, A. Takahashi, T. Yoshida, H. Suzuki, L. C. C. Ambolode II, K. Okazaki, K. Ono, H. Kumigashira, H. Anzai, M. Arita, H. Namatame, M. Taniguchi, D. Ootsuki, K. Sawada, M. Takahashi, T. Mizokawa, Y. Koike, and A. Fujimori,
 arXiv: 1502.03395.

\bibitem{uemura}
Y. J. Uemura, W. J. Kossler, X. H. Yu, J. R. Kempton, H. E. Schone, D. Opie, C. E. Stronach, D. C. Johnston, M. S. Alvarez, and D. P. Goshorn,
 Phys. Rev. Lett. {\bf 59}, 1045 (1987).

\bibitem{hord}
R. Hord, H. Luetkens, G. Pascua, A. Buckow, K. Hofmann, Y. Krockenberger, J. Kurian, H. Maeter, H.-H. Klauss, V. Pomjakushin, A. Suter, B. Albert, and L. Alff,
 Phys. Rev. B {\bf 82}, 180508(R) (2010).

\bibitem{kadono}
R. Kadono, K. Ohishi, A. Koda, W. Higemoto, K. M. Kojima, S. Kuroshima, M. Fujita, and K. Yamada,
 J. Phys. Soc. Jpn. {\bf 72} 2955 (2003).

\bibitem{risdi}
Risdiana, T. Adachi, N. Oki, Y. Koike, T. Suzuki, and I. Watanabe,
 Phys. Rev. B {\bf 82}, 014506 (2010).

\bibitem{asfunction}
In the ZF-$\mu$SR spectra of the as-grown non-SC crystal, effects of small Pr$^{3+}$ moments are masked by effects of large Cu$^{2+}$ moments exhibiting a long-range AF order (see the supplemental materials B).

\bibitem{hlnh}
C. K\"{u}bert, and P. J. Hirschfeld,
 Solid State Commun. {\bf 105}, 459 (1997).

\bibitem{kondo}
As shown in Fig. 3(b) and (c), moreover, the introduction of excess oxygen gives rise to free Cu-spins of LHB of the Cu $3d_{3z^2-r^2}$ orbital in the CuO$_2$ plane just around itself, leading to the occurrence of the Kondo effect.~\cite{adachi}  

\bibitem{mukuda}
for example, see the review paper: H. Mukuda, S. Shimizu, A. Iyo, and Y. Kitaoka,
 J. Phys. Soc. Jpn. {\bf 81}, 011008 (2012).

\bibitem{hiraishi}
M. Hiraishi, R. Kadono, M. Miyazaki, I. Yamauchi, A. Koda, K. M. Kojima, M. Ishikado, S. Wakimoto, and S. Shamoto,
 J. Phys. Soc. Jpn. {\bf 83}, 103707 (2014).

\bibitem{armitage}
N. P. Armitage, F. Ronning, D. H. Lu, C. Kim, A. Damascelli, K. M. Shen, D. L. Feng, H. Eisaki, Z.-X. Shen, P. K. Mang, N. Kaneko, M. Greven, Y. Onose, Y. Taguchi, and Y. Tokura,
 Phys. Rev. Lett. {\bf 88}, 257001 (2002).

\bibitem{matsui}
H. Matsui, K. Terashima, T. Sato, T. Takahashi, S.-C. Wang, H.-B. Yang, H. Ding, T. Uefuji, and K. Yamada,
 Phys. Rev. Lett. {\bf 94}, 047005 (2005).

\bibitem{tsutsumi}
K. Tsutsumi, K. Sato, M. Fujita {\it et al}.,
 unpublished.

\bibitem{kojima}
K. M. Kojima, Y. Krockenberger, I. Yamauchi, M. Miyazaki, M. Hiraishi, A. Koda, R. Kadono, R. Kumai, H. Yamamoto, A. Ikeda, and M. Naito,
 Phys. Rev. B {\bf 89}, 180508(R) (2014).

\end{references}
\end{document}